\documentclass[%
 reprint,
superscriptaddress,
nofootinbib,
 amsmath,amssymb,
 aps,
pra,
floatfix,
]{revtex4-1}
\usepackage[section]{placeins}
\usepackage{graphicx}
\usepackage{dcolumn}
\usepackage{bm}

\usepackage[version=3]{mhchem}
\usepackage{color}
\usepackage{float}
\usepackage{bigstrut}
\usepackage{nicefrac}

\begin{document}

%
\title{Large Resistivity Reduction in Mixed-Valent \ce{CsAuBr3} Under Pressure}
\author{Pavel Naumov}
\affiliation{Max Planck Institute for Chemical Physics of Solids, 01187 Dresden, Germany}
\affiliation{Shubnikov Institute of Crystallography, Russian Academy of Sciences, Moscow 119333, Russia}

\author{Shangxiong Huangfu}
\affiliation{Department of Physics, University of Zurich, CH-8057 Zurich, Switzerland}
\author{Xianxin Wu}
\affiliation{Institute for Theoretical Physics and Astrophysics, University of W\"urzburg, 97074 W\"urzburg, Germany}
\author{Andreas Schilling}
\affiliation{Department of Physics, University of Zurich, CH-8057 Zurich, Switzerland}
\author{Ronny Thomale}
\affiliation{Institute for Theoretical Physics and Astrophysics, University of W\"urzburg, 97074 W\"urzburg, Germany}
\author{Claudia Felser}
\affiliation{Max Planck Institute for Chemical Physics of Solids, 01187 Dresden, Germany}
\affiliation{John A. Paulson School of Engineering and Applied Sciences, Harvard University, Cambridge, Massachusetts 02138, USA}
\author{Sergey Medvedev}
\affiliation{Max Planck Institute for Chemical Physics of Solids, 01187 Dresden, Germany}
\author{Harald O. Jeschke}
\affiliation{Research Institute for Interdisciplinary Science, Okayama University, Okayama 700-8530, Japan}
\author{Fabian O. von Rohr}
\affiliation{Department of Physics, University of Zurich, CH-8057 Zurich, Switzerland}
\affiliation{Department of Chemistry, University of Zurich, CH-8057 Zurich, Switzerland}
\date{\today}
%
\begin{abstract}
We report on high-pressure $p \leq 45$ GPa resistivity measurements on the perovskite-related mixed-valent compound \ce{CsAuBr3}. The compounds high-pressure resistivity can be classified into three regions: For low pressures ($p < 10$ GPa) an insulator to metal transition is observed; between $p= 10$ GPa and 14 GPa the room temperature resistivity goes through a minimum and increases again; above $p = 14$ GPa a semiconducting state is observed. From this pressure up to the highest pressure of $p = 45$ GPa reached in this experiment, the room-temperature resistivity remains nearly constant. We find an extremely large resistivity reduction between ambient pressure and 10 GPa by more than 6 orders of magnitude. This decrease is among the largest reported changes in the resistivity for this narrow pressure regime. We show - by an analysis of the electronic band structure evolution of this material - that the large change in resistivity under pressure in not caused by a crossing of the bands at the Fermi level. We find that it instead stems from two bands that are pinned at the Fermi level and that are moving towards one another as a consequence of the mixed-valent to single-valent transition. This mechanism appears to be especially effective for the rapid buildup of the density of states at the Fermi level.
\end{abstract} 
\maketitle
%
%
\section{Introduction}
Mixed-valent compounds are a class of materials that display disproportionated valences for one or more elements \cite{Rao}. The crucial prerequisite for this is a variable valency of at least one of the elements in the compound. Most often this is observed for compounds of transition metals, lanthanides, or elements with stable $s^2$ and $s^0$ electronic configurations, such as Sn, Sb, Tl, and Bi. Generally, mixed-valent compounds are categorized in three different classes. In a class I system, the ions of differing valences are in sites of different symmetry and ligand field strength, due to their different coordination and bonding properties (e.g. \ce{Fe3O4} spinel) \cite{Fe3O4}. Class II systems are an intermediate state, where the elements in different oxidation states occupy similar sites that are crystallographically distinguishable (e.g. \ce{Ni$_{1-x}$Li$_x$O}) \cite{NiO}. Meanwhile, for class III mixed-valent compounds, the sites that the mixed-valent ions occupy are identical. In this case, the electron hopping is favorable and the electrons are delocalized. This is caused by the the comproportionation of the mixed-valent elements into an on average unfavorable oxidation state. By their very nature of being mixed valent, the cations in class III materials will contribute, on average, an intermediate number of electrons to the band, filling it only partially, while in class I and II materials the bands will be either completely empty or completely filled.

The delocalized electrons in a mixed-valent compound may lead to non-trivial electronic properties. These materials have, therefore, been found to host a variety of electronic instabilities: The most prominent examples for mixed-valent superconductors are the spinel \ce{LiTi2O4}, with a critical temperature of $T_{\rm c} \approx$~12~K \cite{LiTi2O4} and the mixed-valent bismuthate \ce{Ba$_{1-x}$K$_x$BiO3} with a critical temperature of $T_{\rm c} \approx$~30~K \cite{BiO3}. Other electronic properties related to a mixed valence are the Peierls instability in Krogmann's salt \cite{Krog}, the high electrical mobility in the inverse spinel \ce{MoFe2O4} \cite{MoFe2O4}, and the evolution of atypical metallic conductivity in \ce{La_{1-x}Sr_xCoO3} \cite{LaSrCoO3}.

Systems like the mixed-valent bismuthate \ce{Ba$_{1-x}$K$_x$BiO3} are of great interest since they can be tuned through a transition from a class I/class II system to a class III material. Similar systems, e.g. \ce{RbTlCl3}, have very recently been discussed as new potential hosts for non-trivial electronic properties \cite{RbTlCl3,RbTlCl3_2}. The structurally and electronically related compounds \ce{CsAu$X$3} with \textit{X} = Cl, Br, and I crystallize in the space group \textit{I}4/\textit{mmm} in a distorted variation of the perovskite structure \cite{CsAuX3_struct,CsAuX3_2,CsAuX3_struct,Kojima1994}. The gold atoms in \ce{CsAu$X$3} are mixed valent with Au(I) and Au(III) centers, which occupy two different crystallographic sites at ambient conditions \cite{Kitagawa,Liu2000}. This is because Au(II) is an extremely rare oxidation state for gold in solid-state compounds and is only observed for a few mono-nuclear gold labile species under ambient conditions~\cite{Elder1997,Preiss2017}. The three-dimensional metal-halogen frameworks in \ce{CsAuX3} are formed by elongated octahedra with Au(III) as the central atom and compressed octahedra around Au(I); therefore, the chemical formula of these compounds is sometimes written as \ce{Cs2Au2X6}, emphasizing the different nature of the two gold centers. The members of the \ce{CsAuX3} mixed-valence perovskites undergo pressure-induced structural phase transitions at $p \approx$ 11, 9, and 5 GPa, respectively \cite{CsAuX3_struct,CsAuI3_1,CsAuI3_2}. These transitions are associated with a transition from the mixed valency Au(I)/Au(III) to a single valent material with Au(II) at the center of the octahedra \cite{Reis12,Kojima1994}. These transitions are analogous to what has been observed in \ce{BaBiO3}. In the bismuthates, the strong coupling between the charge disproportionation of the metal ion and the lattice deformation stabilizes a commensurate charge-density wave (CDW) \cite{BiO3_2,BiO3_3}. Resistivity measurements on \ce{CsAuCl3}, \ce{CsAuBr3} and \ce{CsAuI3}, up to a pressure $p \approx$ 8 GPa, have revealed an insulator-to-metal transition accompanied with the structural transitions \cite{Kitagawa_2,Kojima1994}. 

Here, we report on the electronic transport properties of \ce{CsAuBr3} under applied high pressures up to $p =$ 45 GPa. We show that this compound undergoes a drastic, continuous change of its electronic properties from an insulator (at ambient pressure) to a strange metallic state ($p \approx$ 10 GPa), and back to a semiconductor ($p \approx$ 14 GPa). We observe one of the largest resistivity decreases at room-temperature within the first 10 GPa that has been reported. We analyze this observation by a detailed density functional theory investigation of the electronic structure evolution with pressure.

\section{Experiment and theory}
Large single crystals of \ce{CsAuBr3}, up to dimensions of $V \approx$ 6 x 4 x 0.8 mm$^3$, were grown by a self-flux method adapted from reference ~\cite{Chan}, using stoichiometric amounts of CsBr (purity 99.999 \%, anhydrous), liquid bromine (purity 99.998 \%), and gold powder (purity 99.99 \%). All chemicals were placed in a corundum crucible inside a quartz tube. The quartz tube was cooled by immersion in liquid nitrogen to avoid evaporation of bromine. The quartz tube was sealed under vacuum and slowly heated to 650 $^\circ$C, held at this temperature for 48 h, and slow cooled with 1 $^\circ$C/h to room temperature. The crystal structure and phase purity of the samples were investigated by powder x-ray diffraction (PXRD) measurements on a STOE STADIP diffractometer with Cu K$_{\alpha 1}$ radiation. The PXRD pattern was collected in the $2 \Theta$ range of 8-90$^\circ$ with a scan rate of 0.25$^\circ$/min. The \textit{Le Bail} fit was performed using the FULLPROF program package.

A non-magnetic diamond anvil cell was used for the high-pressure measurements under $p$ values of up to 45~GPa \cite{S_1,S_2}. Single crystals, as grown by the procedure described above, were used for the high-pressure experiments. A cubic BN/epoxy mixture was used for the insulating gaskets and platinum foil was employed for the electrical leads. No pressure transmitting medium was used. Electrical resistivity was measured  in a customary cryogenic setup using the dc current in a van der Pauw geometry (lowest achievable temperature is $T =$~1.5~K). The high pressure Raman spectra were recorded using a customary micro-Raman spectrometer with a HeNe laser as the excitation source and a single-grating spectrograph with a resolution of 1~cm$^{-1}$. Raman scattering was calibrated using Ne lines with an uncertainty of 1~cm$^{-1}$. The pressure was determined using the ruby scale \cite{Mao1986} by measuring the luminescence from the small chips of ruby placed in contact with the sample.

A projector augmented wave basis as implemented in VASP~\cite{Kresse1993,Kresse1996,Kresse1996-2} for the prediction of high pressure crystal structures of \ce{CsAuBr3} was used. A generalized gradient approximation (GGA)~\cite{Perdew1996} exchange correlation functional was used and the electronic structure was converged using $9\times 9\times 9$ $k$ points in the Brillouin zone. In the relaxation, the plane wave cutoff energy is 600 eV and forces are minimized to less than 0.01~eV/{\AA} in the relaxation. The precise electronic structures were determined, using the fully relativistic version of the all electron full potential local orbital (FPLO)~\cite{Koepernik1999} basis and GGA exchange correlation functional. All calculations were converged using $24\times 24\times 24$ $k$ points.
\section{Results and Discussion}

\begin{figure}
	\centering
	\includegraphics[width=0.95\linewidth]{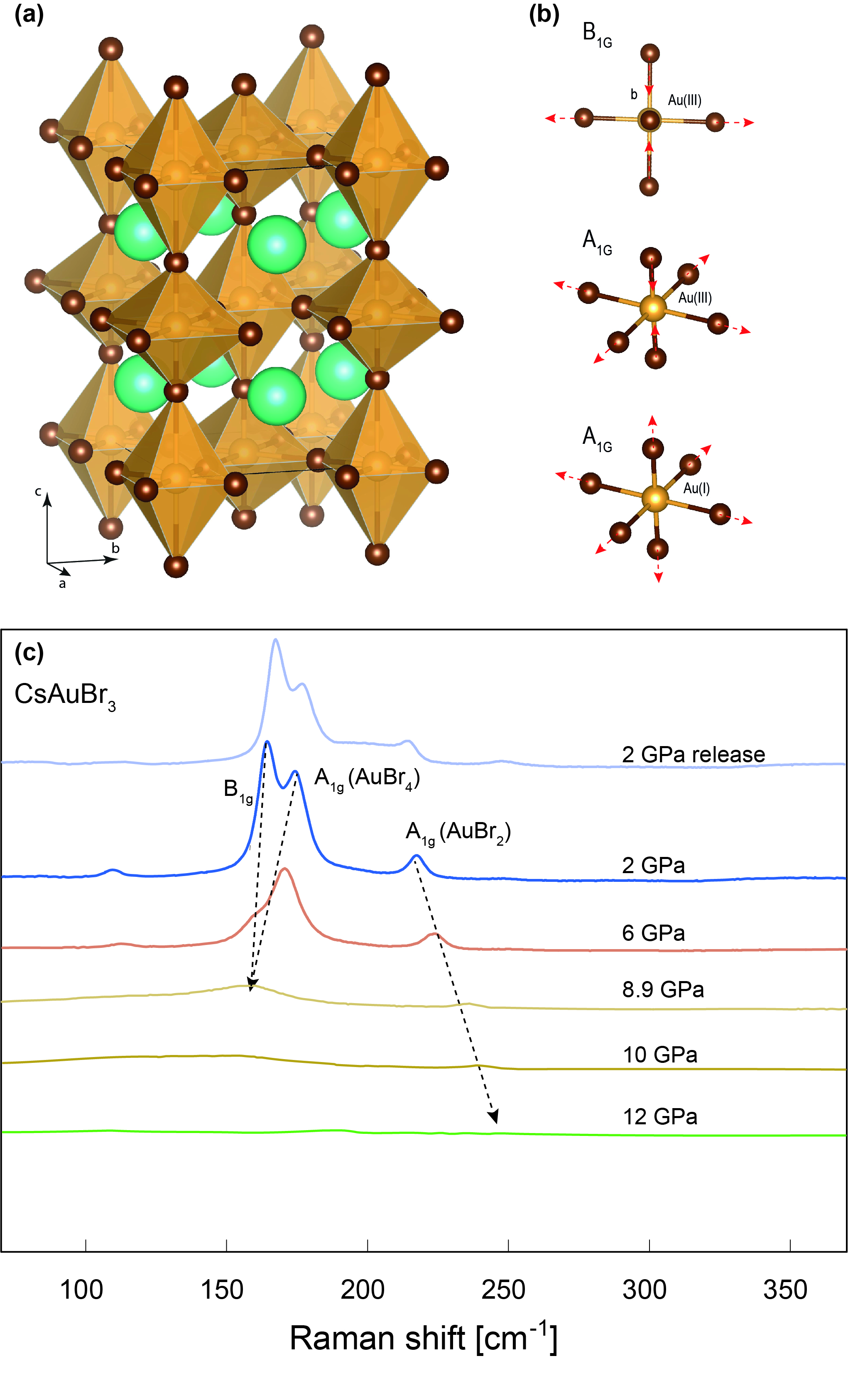}
	\caption{(a) Crystal structure of \ce{CsAuBr3} (b) Raman modes of the gold octahedra. (c)  Raman spectra for pressures ranging from $p = 2$ GPa to 12 GPa. For higher pressures, no peaks in the Raman spectra were observed. The Raman spectra after the released pressure, at 2 GPa is shown on top, indicating full reversibility. The spectra at 8.9, 10, and 12 GPa are multiplied by a factor of 4 and 20, respectively, for better comparability.}
	\label{fig:raman}
\end{figure}

The crystal structure of \ce{CsAuBr3} is shown in Figure~\ref{fig:raman}(a). This material is a mixed-valent compound at ambient pressure; its stoichiometric formula can also be written as \ce{Cs2(AuBr4)(AuBr2)} or \ce{Cs2Au2Br6} in order to emphasize the difference of the respective Au-centers. The planar \ce{AuBr4-} and the linear \ce{AuBr2-} anions lead to the distorted octahedral coordination of the Au-centers. Due to these inequivalent Au-centers, the compound crystallizes in a distorted version of the perovskite structure. The corresponding vibrations of these distorted octahedra are schematically shown for the structures of the \ce{AuBr4-} and \ce{AuBr2-} anions in the mixed-valent state, and the schematic structure of the single-valent state is depicted in Figure~\ref{fig:raman}(b).

In Figure~\ref{fig:raman}(c), we show the measured Raman spectra of \ce{CsAuBr3} for the pressures $p = 2.0$, 6.0, 8.9, 10.0, and 12.0 GPa. The shift positions of the Raman spectrum at $p = 2$ GPa is in good agreement with earlier values \cite{Raman,Raman2}. The Raman spectrum of \ce{CsAuBr3} includes three Au-Br stretching modes at 167 cm$^{-1}$ (B$_{\rm 1g}$, Au(III)Br$_4$), at 179 cm$^{-1}$ (A$_{\rm 1g}$ Au(III)Br$_4$), and at 220 cm$^{-1}$ (A$_{1g}$ Au(I)Br$_{2}$), respectively. In the single-valent state, the halogen atoms are located at the central position between the two neighboring gold sites, leading to a symmetric \ce{AuBr6} coordination, and thus the gold-halogen stretching Raman bands are deactivated. 

For our \ce{CsAuBr3} sample we observe the Au(III) modes to soften, while the frequency of the Au(I) mode increases with increasing pressure. At a pressure of $p = 7.5$ GPa, all active Raman modes have disappeared or their intensities are too low for detection. This observation is in good agreement with the realization of a single-valent state. A further pressure increase does not lead to significant changes in the Raman spectra, but mainly to a weakening of any Raman shifts. The Raman spectra recorded after the pressure release matches the spectra of the unpressurized sample, which indicates that the pressure-induced structural changes are reversible.

\begin{figure}
	\centering
	\includegraphics[width=\linewidth]{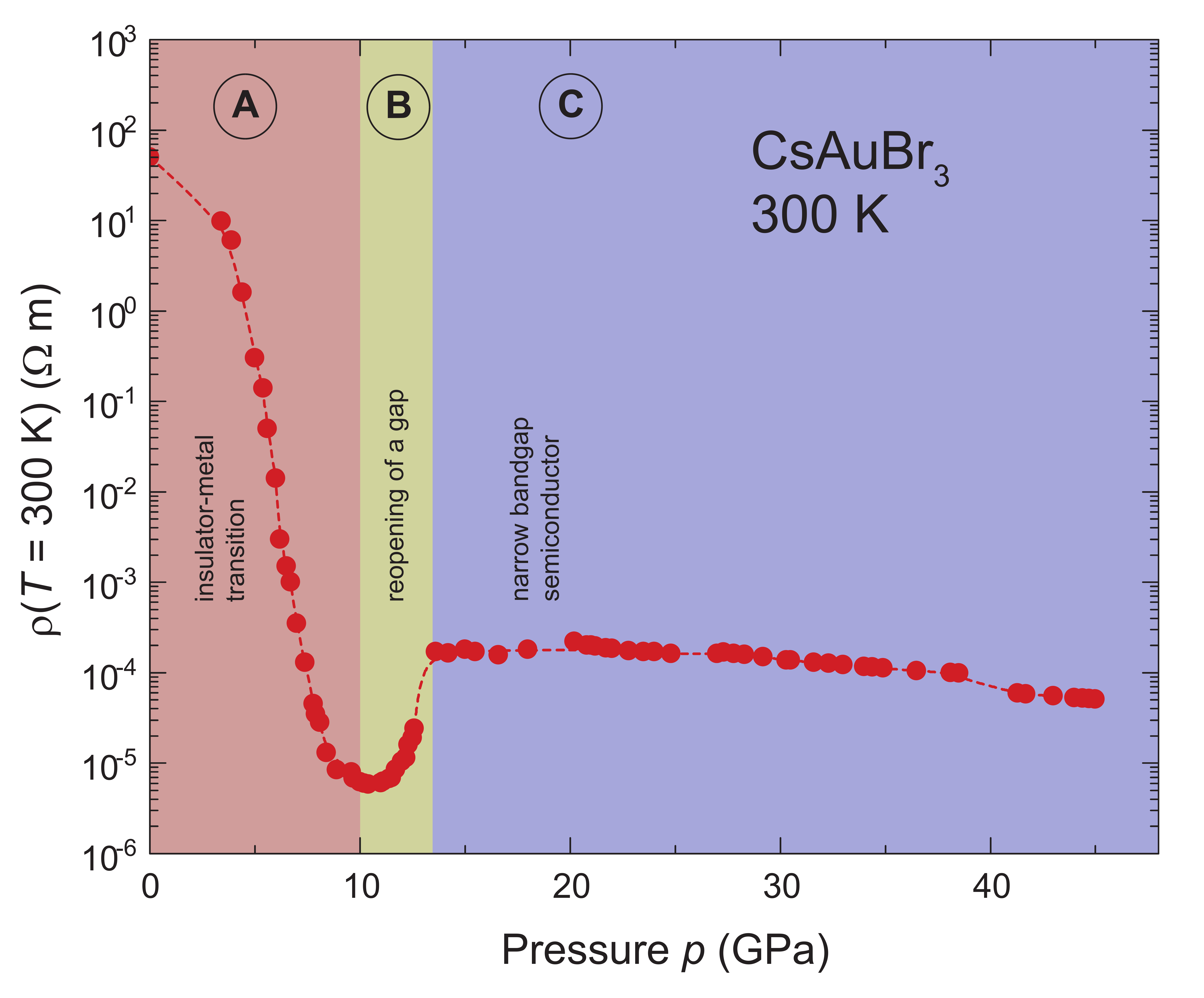}
	\caption{Evolution of the resistivity of \ce{CsAuBr3} at 300 K in a pressure range up to $p = 45$ GPa. The highlighted region A corresponds to an insulator to metal transition. In region B an electronic gap is reopened. In region C corresponds to a robust semiconducting state at high-pressures.}
	\label{fig:300K}
\end{figure}

In Figure \ref{fig:300K}, we show the resistivity at room temperature of \ce{CsAuBr3} as a function of the quasi-hydrostatic pressure, up to a pressure of $p = 45$ GPa. Our resistivity measurement at room temperature shows that under pressures up to 10 GPa, the resistivity of \ce{CsAuBr3} decreases by several orders of magnitude. With a further pressure increase, the resistivity reaches a minimum at around $p = 10$ GPa, and then increases again to reach a saturation value at pressures larger than $p = 14$ GPa. The resistivity evolution under pressure can be divided into three distinct regions: 

The highlighted region A corresponds to an insulator to metal transition. Between ambient pressure and a pressure of $p \approx 10$ GPa, the resistivity decreases drastically from $\rho_{\rm 300 K} \approx 48 \ \Omega \rm{m}$ to $5.8 \cdot 10^{-6} \ \Omega \rm{m}$. Going from an overall insulating resistivity to the resistivity comparable to a good conductor such as a pure metal, Region B corresponds to an insulator to metal transition. Around 10 GPa, the room temperature resistivity goes through a minimum and increases again between $p = 12$ to 14 GPa to a value of $\rho_{\rm 300 K} \approx 1.7 \cdot 10^{-4} \ \Omega \rm{m}$. Region C corresponds to a semiconducting state, where an electronic bandgap has reopened. Above a pressure of $p > 14$ GPa, the room-temperature resistivity remains nearly constant up to the  maximum pressure of $p = 45$ GPa reached in this experiment.

\begin{figure}
	\centering
	\includegraphics[width=0.95\linewidth]{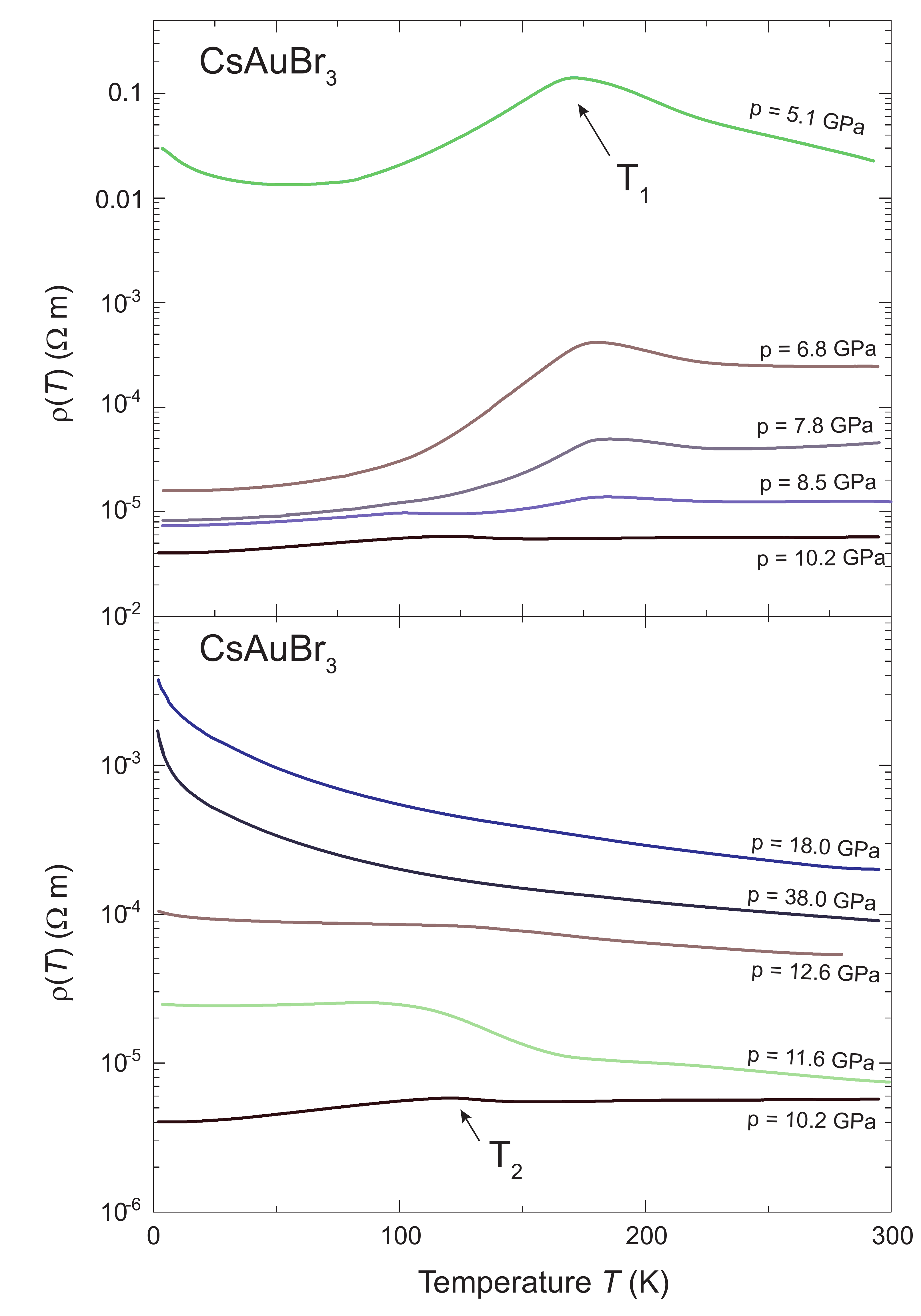}
	\caption{Temperature-dependent resistivity of \ce{CsAuBr3} in a temperature range between $T = 300$ K and 1.5 K for pressures of $p = 5.1$, 6.8, 7.8, 8.5, 10.2, 11.6, 12.6, 18.0, 38.0 GPa. The anomalies in the resistivity are marked with $T_1$ and $T_2$.}
	\label{fig:rhoT}
\end{figure}

The resistivity decrease of \ce{CsAuBr3} within the first 10 GPa is one of the largest reported in the literature. For comparison, the compound \ce{VO2}, which is a prototype material for studying metal-to-insulator transitions displays a resistivity change of only approximately one order of magnitude within the first 10 GPa \cite{VO2}. Black phosphorus \cite{black_P1}, which undergoes a semiconductor to metal transition under pressure, associated with two continuous structural transitions and \ce{BaFe2Se3}, a spin-chain material that displays pressure-induced superconductivity have both changes in resistivity of only approximately 3 orders of magnitude \cite{black_P1,black_P2,BaFe2Se3}.

SmTe and SmSe have been reported to have resistivity changes comparable to the one reported here for \ce{CsAuBr3} within 10 GPa \cite{SmTe}. These rare earth monochalcogenides crystallize in the NaCl-type structure and are semiconducting or metallic depending on the valence state of the rare earth ion. Upon the application of pressure, these compounds display a transition from Sm(II) to Sm(III) \cite{SmTe_2}. Similar to the case of \ce{CsAuBr3}, the change in resistivity is associated with a change in valence state of an element.   

We have performed temperature-dependent resistivity measurements for pressures between $p =  5.1$ to 38.0 GPa and for temperatures of $T = 300$~K to 1.5 K; these are shown in Figure~\ref{fig:rhoT}. For the pressure regime below $p \approx 10$ GPa, these measurements further support the gradual transition from an insulating state to metallic behavior. This can be observed by the overall trend change from a increasing resistivity (with decreasing temperature) at a pressure of $p = 5.1$ GPa to a decreasing one, e.g. at $p = 6.9$ GPa. The temperature-dependent resistivities in the metallic states are never Fermi-liquid like. 

The gradual decrease from semiconductor to metallic behavior of the resistivity under pressure is accompanied by an anomaly at $T_1 = 170$~K. This anomaly weakens for increasing pressures, but it remains observable for all samples (see, mark in Figure~\ref{fig:rhoT}(a)) At the pressures of $p =$ 8.5 GPa and 10.2 GPa, where an almost completely metallic dependence of the resistivity is reached, a second anomaly at $T_2 \approx 100$ K is observed. Furthermore, the very small RRR values indicate that either a very small fraction of electrons is only participating in the electronic transport, or that the transport is strongly defect dominated (compare, e.g., references \cite{Jia_defect,GeSe}). With the further increase of pressure to $p > 14$ GPa, the resistivity increases with decreasing temperature. The overall behavior of the resistance becomes semiconducting. 

The increase of the room-temperature resistivity (Figure~\ref{fig:300K}) for pressures above 14 GPa can therefore be directly connected to the reopening of an electronic gap at the Fermi level. We find this high pressure semiconducting state to be remarkably robust, with almost no further change up to a pressure of $p =$ 38 GPa. Both the occurrence of the electronic anomalies and the reopening of a gap are in good agreement with earlier pressure resistivity, infrared, and Raman spectroscopy measurements on \ce{CsAuI3} \cite{Kojima1994,CsAuI3_1}. \ce{CsAuBr3} is not found to become superconducting for any of the pressures, despite the very low resistivities and high conductivities observed. Investigation of the origin of the anomalies at $T_1$ and $T_2$, for example by assessing the relevance of strong electronic correlations within a combination of density functional theory and dynamical mean field theory or by probing possible magnetic tendencies of Au$^{2+}$, will be promising future research directions.

\begin{figure}
	\includegraphics[width=0.9\linewidth]{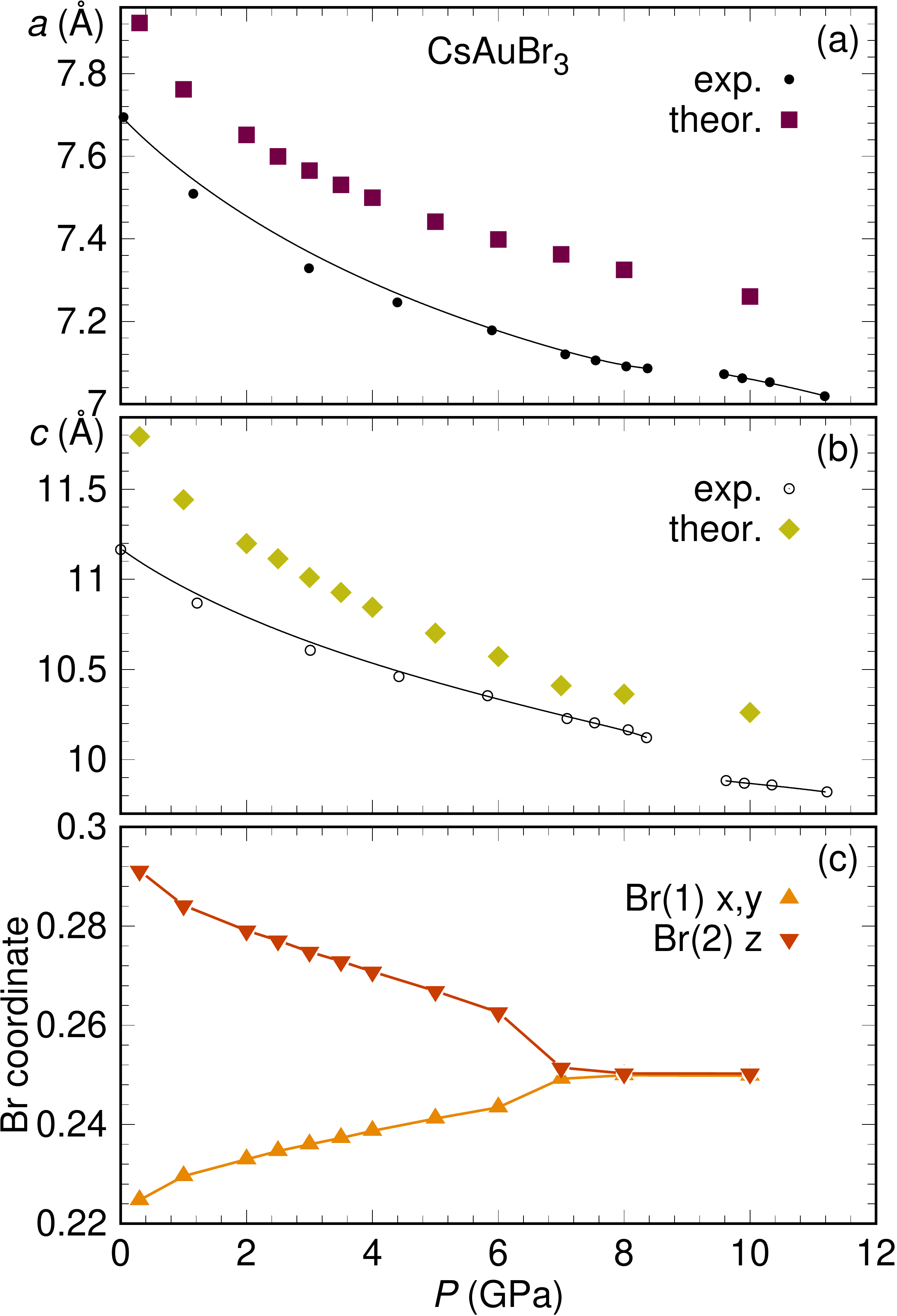}
	\caption{Pressure evolution of structural parameters of \ce{CsAuBr3}, comparing experimental data of reference~\protect\onlinecite{Kojima1994} and theoretical predictions.}
	\label{fig:structure}
\end{figure}

\begin{figure}
	\includegraphics[width=0.95\linewidth]{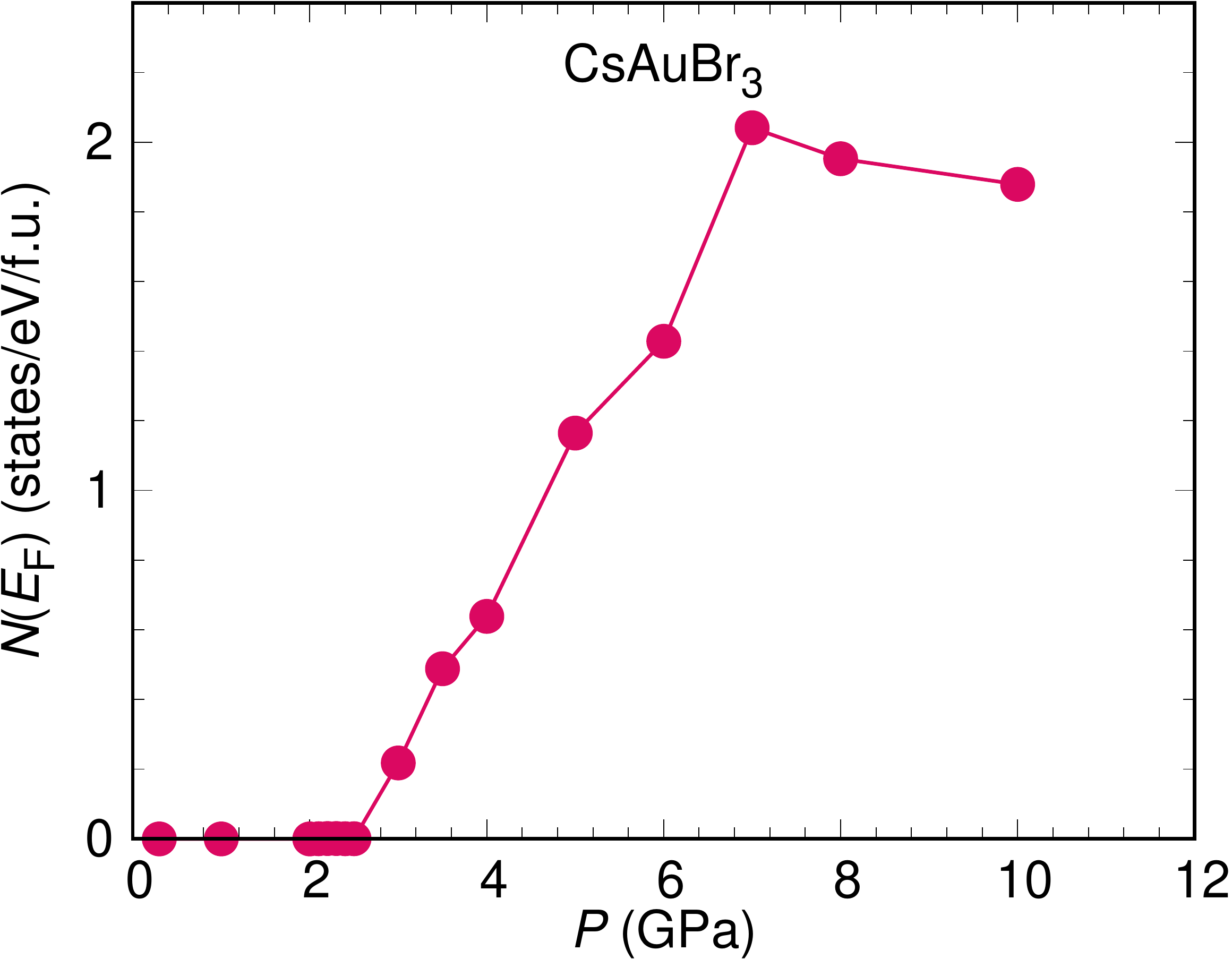}
	\caption{Calculated pressure evolution of the density of states at the Fermi level of \ce{CsAuBr3}.}
	\label{fig:nef}
\end{figure}

\begin{figure}
	\includegraphics[width=0.95\linewidth]{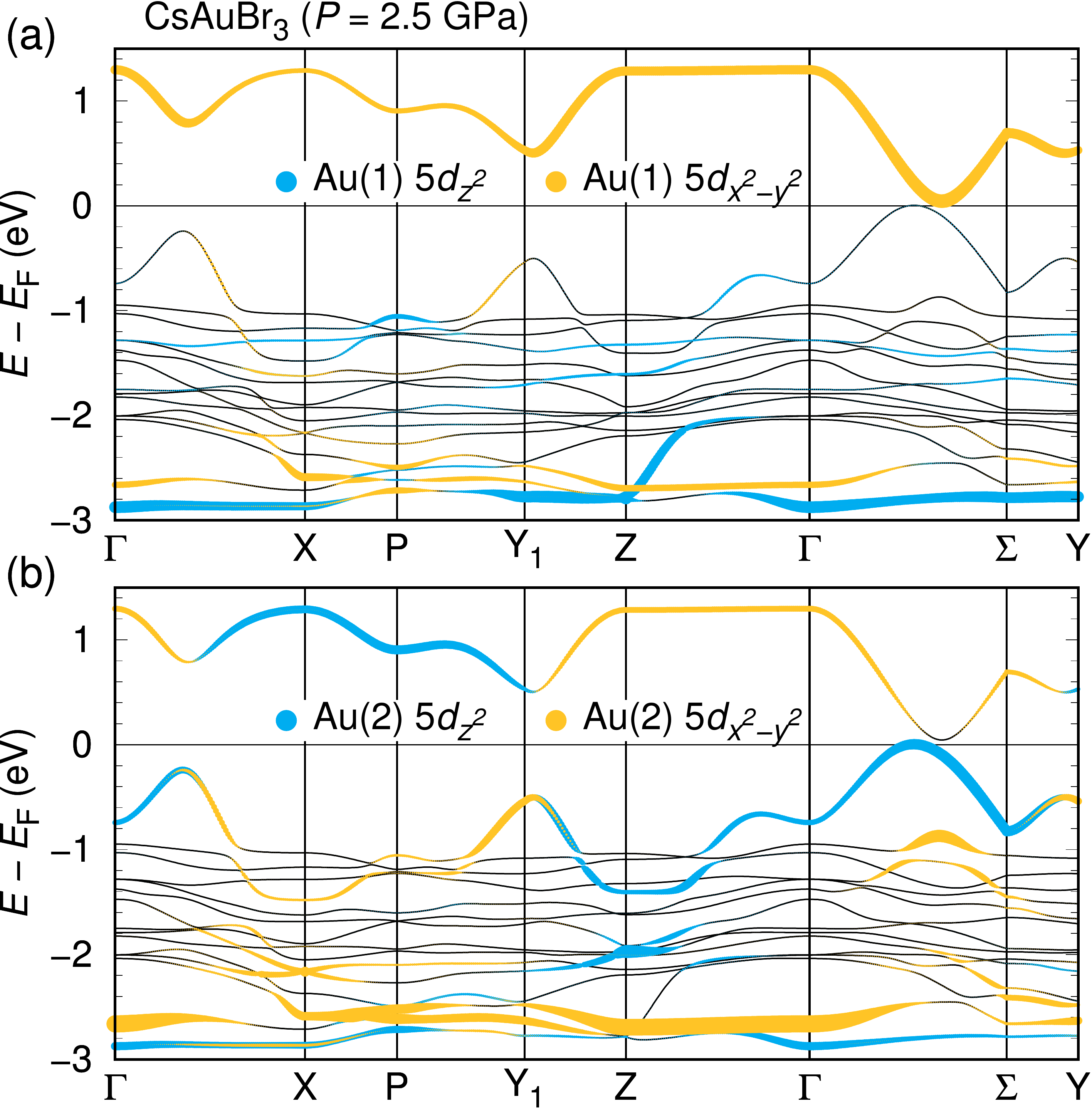}
	\caption{Orbital weights of the two inequivalent Au ions in the unit cell of \ce{CsAuBr3}.}
	\label{fig:weights}
\end{figure}

For the theoretical analysis of the electronic structure of \ce{CsAuBr3} under pressure up to a pressure of $p =$ 10 GPa, the available crystallographic data from reference~\cite{Kojima1994} were used. \ce{CsAuBr3} crystallizes in the tetragonal space group $I4/mmm$ with $a$ and $c$ lattice parameters. Cs, Au(1) and Au(2) are in fixed positions $(0,\nicefrac{1}{2},\nicefrac{1}{4}$ ($4d$), $(0,0,0)$ ($2a$) and $(0,0,\nicefrac{1}{2})$ ($2b$), respectively. Br(1) and Br(2) are in $(x,x,0)$ ($8h$) and $(0,0,z)$ ($4e$) positions, respectively, so that the space group has two more free parameters, $x$ and $z$. We find that simple relaxation of these two internal positions for the given experimental lattice parameters severely overestimates the metallicity of \ce{CsAuBr3}. Such a calculation predicts an insulator only in a negligible pressure interval. Therefore, we have used a projector augmented wave basis to predict the full structures at various pressures. The results are shown in Figure~\ref{fig:structure}. 

As expected, the lattice parameters are slightly overestimated by this approach, because the GGA functional has a tendency to underestimate the chemical bonding strength in materials. The cell parameters and their trend with a function of increasing pressure are, however, in good agreement with the experimental values. We also tested the use of the local density approximation (LDA), but the results of GGA functional are in better agreement with the experimental parameters for structure and electronic properties. The results of the structure optimization, specifically the cell parameters $a$ and $c$, for \ce{CsAuBr3} in comparison with the experimental values are shown in Figure~\ref{fig:structure}(a)-(b).

The pressure evolution of the Br(1) and Br(2) positions, shown in Figure~\ref{fig:structure}~(c), can be interpreted as an order parameter for the mixed-valent to single-valent phase transition of \ce{CsAuBr3} under pressure. When the Br(1) and Br(2) positions become equivalent then the single-valent state is reached. We find the pressure induced phase transition from mixed-valent, $I4/mmm$, to the single valent state, $Pm\bar{3}m$, to occur between 7 and 8~GPa in our calculations. The higher symmetry is realized when both Br(1) and Br(2) are on the  $x=y=\nicefrac{1}{4}$ and $z=\nicefrac{1}{4}$ positions, respectively. Experimentally, the structural phase transition is reported at about 9~GPa~\cite{Kojima1994}. Note that our structure predictions do not allow us to resolve a tetragonal to tetragonal phase transition before the material becomes cubic. The bands calculated for the experimental $P=8.1$~GPa structure determined by Sakata {\it et al.}~\cite{Sakata2004} are very similar to figure 7~(b), calculated for the structure predicted at $P=6$~GPa.

In Figure~\ref{fig:nef}, we show the pressure evolution of the density of states at the Fermi level $N(E_{\rm F})$, calculated with fully relativistic GGA. We can distinguish three pressure regions: An insulating region up to $p=2.5$~GPa, a sharp increase of $N(E_{\rm F})$ up to $p=7$~GPa, and a gradual decrease of $N(E_{\rm F})$ for higher pressures. We can identify these three regions with the pressure evolution of the conductivity (Figure~\ref{fig:rhoT}); the conductivity decreases only by a factor of five between 0~GPa and 3.4~GPa, corresponding to the gradual closing of the charge gap; then it decreases by six orders of magnitude between 3.4~GPa and about 8~GPa, corresponding to the rapid increase of $N(E_{\rm F})$; finally, after going through a minimum it increases a bit, corresponding to the slow decrease of $N(E_{\rm F})$.

\begin{figure}
	\includegraphics[width=0.7\linewidth]{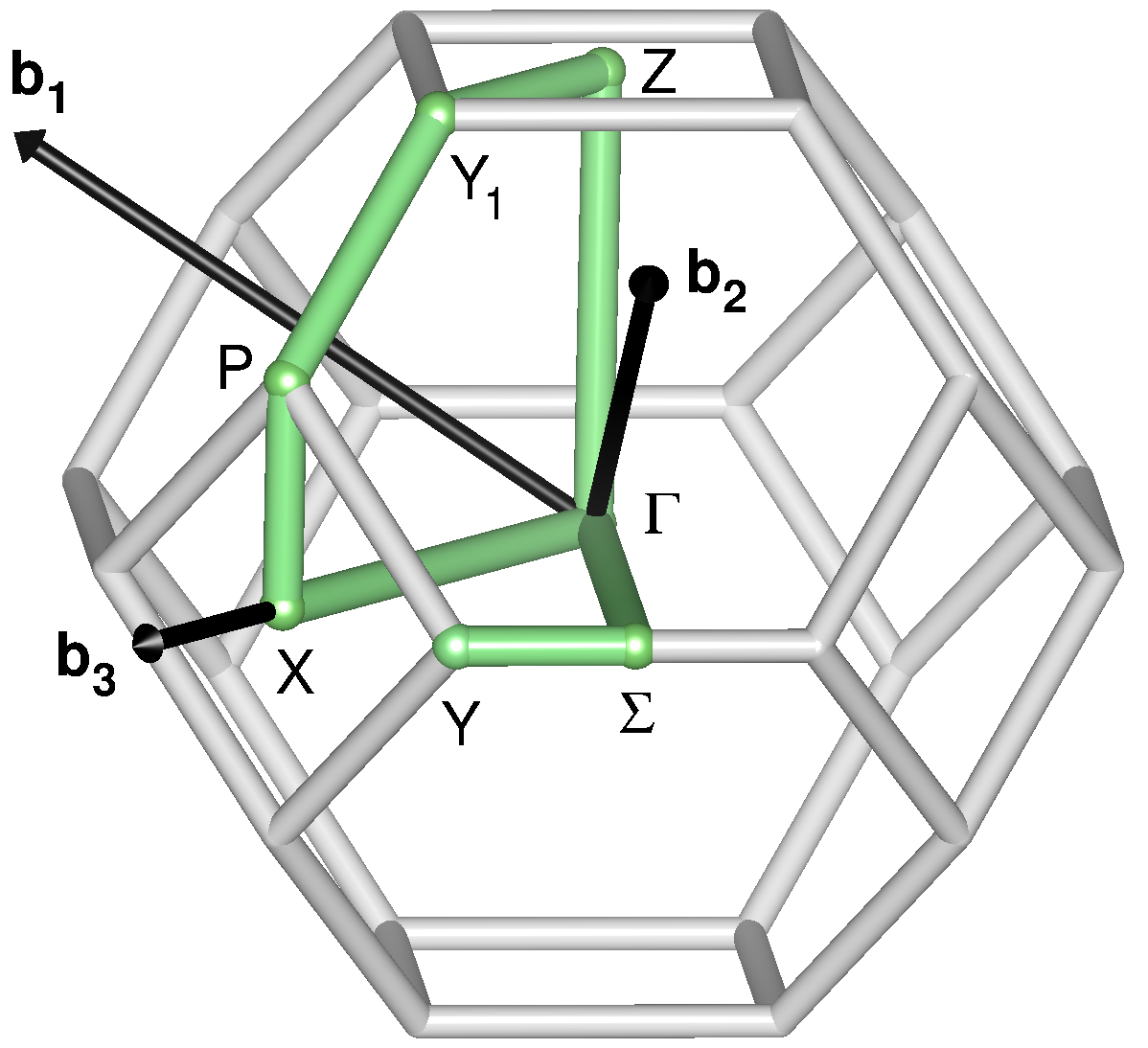}
	\caption{Brillouin zone, marked in grey, and the path chosen, marked in green, for the electronic band structure calculations in this work.}
	\label{fig:Brillouin}
\end{figure}

\begin{figure}
	\includegraphics[width=0.85\linewidth]{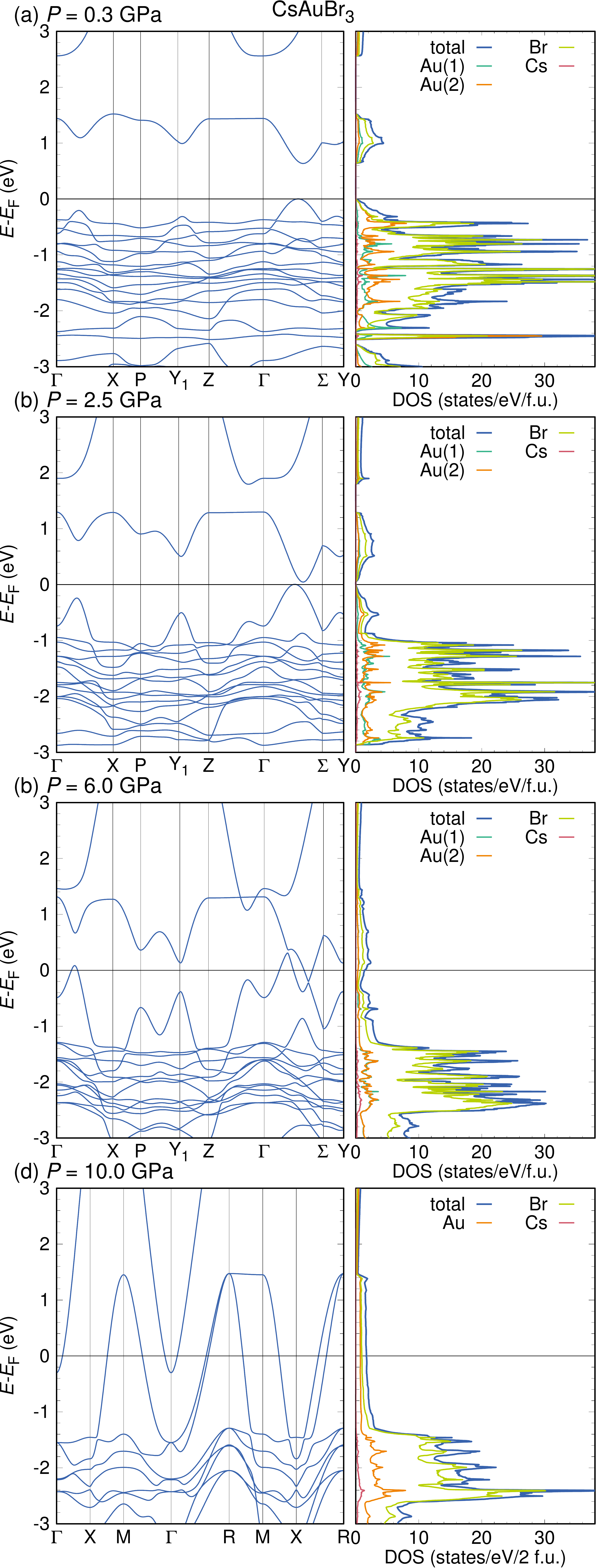}
	\caption{Electronic bands structures and densities of states of \ce{CsAuBr3} at selected pressures. (a) to (c) are calculated using $I4/mmm$ space group for one \ce{CsAuBr3} formula unit, (d) with $Pm\bar{3}m$ space group for one \ce{CsAuBr3} formula unit.}
	\label{fig:bands}
\end{figure}

Figure~\ref{fig:weights} clarifies the character of the bands above and below the Fermi level that are brought closer under the influence of pressure until they overlap. At $p=2.5$~GPa, near the critical pressure for the metal insulator transition as determined by DFT, the valence band has predominantly Au(2) $5d_{z^2}$ character, and the conduction band is mostly of Au(1) $5d_{x^2-y^2}$ character. The high symmetry points in the Brillouin zone are~\cite{Setyawan2010} $X=(0,0,\nicefrac{1}{2})$, $P=(\nicefrac{1}{4},\nicefrac{1}{4},\nicefrac{1}{4})$, $Y_1=(\nicefrac{1}{2},\nicefrac{1}{2},-\zeta)$, $Z=(\nicefrac{1}{2},\nicefrac{1}{2},-\nicefrac{1}{2})$, $\Sigma=(\eta,-\eta,\eta)$, $Y=(\zeta,-\zeta,\nicefrac{1}{2})$ with $\eta=\frac{1}{4}(1+\frac{a^2}{c^2})$, $\zeta=\frac{a^2}{2c^2}$. The Brillouin zone and the path chosen for Figures~\ref{fig:weights} and \ref{fig:bands} are depicted in Figure~\ref{fig:Brillouin}.

The insulator-metal phase transition and the subsequent rapid increase of the density of states at the Fermi level can be understood from the octahedral crystal field of the two inequivalent gold atoms. The Au(1) octahedron is enlarged along $z$, leading to higher $5d_{x^2-y^2}$ energy compared to $5d_{z^2}$, and thus an unoccupied  $5d_{x^2-y^2}$ state. The Au(2) octahedron is compressed along $z$, leading to higher $5d_{z^2}$ energy and therefore a partly unoccupied $5d_{z^2}$ state. The increase of pressure lowers the distortion of both octahedra and thus reduces the crystal field splitting. When the Au(1) $5d_{x^2-y^2}$ and the Au(2) $5d_{z^2}$ meet, the gap closes; the subsequent increase of $N(E_{\rm F})$ is particularly fast because pressure acts on both occupied and unoccupied bands which leads to a strong build-up of density of states at the Fermi level. This fast increase of $N(E_{\rm F})$ within less than $\Delta p =$ 5~GPa corresponds to the steep drop in resistivity by more than six orders of magnitude in a similar pressure window. When the material reaches the single valent state at 7~GPa, the crystal field splitting between $5d_{x^2-y^2}$ and $5d_{z^2}$ reaches zero, which brings the $N(E_{\rm F})$ growth mechanism to a halt. Further increase of pressure leads to the usual band width increase, leading to a slow $N(E_{\rm F})$ decrease and consequently a resistivity increase. Figure~\ref{fig:bands} shows the bands and density of states for a few sample pressures. The $p=10$~GPa bands are shown with the usual high symmetry points of the cubic $Pm\bar{3}m$ space group~\cite{Setyawan2010}. 

\section{Conclusion}

In summary, we have reported on the high-pressure resistivity of the mixed-valent compound \ce{CsAuBr3}, which crystallizes in the space group \textit{I}4/\textit{mmm} in a distorted variation of the perovskite structure. In this compound, the gold atoms are mixed valent with Au(I) and Au(III) centers, which occupy two different crystallographic sites at ambient conditions. Upon application of external pressure, this mixed-valent state transforms into a single-valent state with gold in the uncommon Au(II) oxidation state. The resistivity evolution under pressure can be divided into three distinct regions: For low pressures p $<$ 10 GPa an insulator to metal transition is observed; between $p=$ 10 GPa and 14 GPa the room temperature resistivity goes through a minimum and increases again; above a pressure of p $>$ 14 GPa a semiconducting state emerges, which is related to the reopening of the bandgap. Above a pressure of $p > 14$ GPa the room-temperature resistivity remains nearly constant up to the maximal pressure of $p = 45$ GPa reached in this experiment.

We find the insulator to metal transition for small pressures p $<$ 10 GPa to be accompanied by a large change in resistivity of more than 6 orders of magnitude at room temperature. This change in resistivity is among the largest reported resistivity changes in this narrow pressure regime. We showed by an analysis of the electronic structure evolution of this material that the large change in resistivity under pressure in not caused by some crossing of the bands at the Fermi level, but rather by two bands that are pinned at the Fermi level and that are moving towards one another as a consequence of the mixed-valent to single-valent transition. This mechanism appears to be especially effective for the rapid build up of the density of states at the Fermi level. This mechanism leading to a large change in resistivity of \ce{CsAuBr3} may represent a new direction in the study of stress controlled electronic transport. 

\section{Acknowledgements}

We thank Stefan Siegrist and Daniel Schnarwiler for help with the synthesis of the sample, we furthermore thank Jorge Lago for helpful discussions. This work was supported by the Swiss National Science Foundation under Grant No. PZ00P2\_174015 and 20\_175554. The work in W\"urzburg is supported by the German Research Foundation (DFG) through DFG-SFB 1170, project B04 and by the W\"urzburg-Dresden Cluster of Excellence on Complexity and Topology in Quantum Matter -- \textit{ct.qmat} (EXC 2147, project-id 39085490). P.N. acknowledges the support of high-pressure experiments by the Ministry of Science and Higher Education within the State assignment FSRC  "Crystallography and Photonics" RAS.

\end{document}